\documentclass[aps,prl,twocolumn,superscriptaddress]{revtex4-1}

\usepackage{url}                         
\usepackage{subfigure}
\usepackage{epsfig}
\usepackage{amsmath,amssymb}
\usepackage{longtable}
\usepackage{lineno}
\usepackage{soul}

\usepackage{color}

\begin{document}


\title{First Precision Measurement of the Parity Violating Asymmetry in Cold Neutron Capture on $^3$He.}


\author{M.~T.~Gericke}
\email[]{Michael.Gericke@umanitoba.ca}
\affiliation{University of Manitoba}
\author{S.~Bae{\ss}ler}
\affiliation{University of Virginia}
\affiliation{Oak Ridge National Laboratory}
\author{L.~Barr\'on-Palos}
\affiliation{Universidad Nacional Aut\'{o}noma de M\'{e}xico}
\author{N.~Birge}
\affiliation{University of Tennessee}
\author{J.~D.~Bowman}
\affiliation{Oak Ridge National Laboratory}
\author{J.~Calarco}
\affiliation{University of New Hampshire}
\author{V.~Cianciolo}
\affiliation{Oak Ridge National Laboratory}
\author{C.~E.~Coppola}
\affiliation{University of Tennessee}
\author{C.~B.~Crawford}
\affiliation{University of Kentucky}
\author{N.~Fomin}
\affiliation{University of Tennessee}
\author{I.~Garishvili}
\affiliation{University of Tennessee}
\author{G.~L.~Greene}
\affiliation{University of Tennessee}
\affiliation{Oak Ridge National Laboratory}
\author{G.~M.~Hale}
\affiliation{Los Alamos National Laboratory}
\author{J.~Hamblen}
\affiliation{University of Tennessee Chattanooga}
\author{C.~Hayes}
\affiliation{University of Tennessee}
\author{E.~Iverson}
\affiliation{Oak Ridge National Laboratory}
\author{M.~L.~Kabir }
\affiliation{University of Kentucky}
\author{M.~McCrea}
\affiliation{University of Manitoba}
\affiliation{University of Winnipeg}
\author{E.~Plemons}
\affiliation{University of Tennessee}
\author{A.~Ram\'irez-Morales}
\affiliation{Universidad Nacional Aut\'{o}noma de M\'{e}xico}
\author{P.~E.~Mueller}
\affiliation{Oak Ridge National Laboratory}
\author{I.~Novikov}
\affiliation{Western Kentucky University}
\author{S.~Penttila}
\affiliation{Oak Ridge National Laboratory}
\author{E.~M.~Scott}
\affiliation{University of Tennessee}
\author{J.~Watts}
\affiliation{University of Tennessee Chattanooga}
\author{C.~Wickersham}
\affiliation{University of Tennessee Chattanooga}

\collaboration{The n3He Collaboration}
\noaffiliation

\date{\today}

\begin{abstract}
We report the first precision measurement of the parity-violating asymmetry in the direction of proton emission with respect to the neutron spin, in the reaction $^{3}\mathrm{He}(\mathrm{n},\mathrm{p})^{3}\mathrm{H}$, using the capture of polarized cold neutrons in an unpolarized active $^3\rm{He}$ target. The asymmetry is a result of the weak interaction between nucleons, which remains one of the most poorly understood aspects of electro-weak theory. The measurement provides an important benchmark for modern effective field theory (EFT) calculations. Measurements like this are necessary to determine the spin-isospin structure of the hadronic weak interaction. Our asymmetry result is $A_{PV} = \left( 1.58 \pm 0.97 ~\mathrm{(stat)} \pm 0.24~\mathrm{(sys)}\right)\times10^{-8}$, which has the smallest uncertainty of any parity-violating asymmetry measurement so far.

\end{abstract}


\maketitle

\section{Introduction}

The electroweak (EW) component of the Standard Model (SM) describes the weak couplings of W and Z gauge bosons to quarks and therefore, in principle, the hadronic weak interaction (HWI). In nuclei, the HWI causes parity-violating (PV) admixtures in nuclear wave functions and produces small, but observable, PV spin-momentum correlations, photon circular polarizations and anapole moments. However, the computational difficulties associated with nonperturbative QCD dynamics currently preclude first-principles calculations of hadronic PV observables. As a result, the HWI is the least well understood sector of the Standard Model. The most ambitious effort to carry out a QCD calculation on the lattice has been that of Wassem~\cite{Wasem}.

Desplanques, Donoghue, and Holstein (DDH)~\cite{DDH} introduced a physically motivated meson-exchange potential model. The resulting PV nucleon-nucleon potential is a sum over the 6 parity-odd, time-reversal-even, rotationally invariant operators that can be constructed from the spin, isospin, momenta, and coordinates of the interacting nucleons and 6 meson-exchange coupling constants. Each operator has a Yukawa dependence on the separation of the nucleons with range determined by the mass of the exchanged meson, $m_{\pi}$, $m_{\rho}$, and $m_{\omega}$. The six floating coupling constants ($h^{1}_{\pi}$, $h^{0}_{\rho}$, $h^{1}_{\rho}$, $h^{2}_{\rho}$, $h^{0}_{\omega}$, and $h^{1}_{\omega}$) are labeled by the meson exchanged and of the total isospin change ($\Delta I$) of the interacting pair. The pion exchange potential is unique in having a longer range than the other terms. DDH give reasonable ranges and best values for these 6 couplings~\cite{DDH}. The PV observables for each system are calculated as sums of matrix elements of the PV potential terms between nuclear states, multiplied by the coupling constants (see Eqn.~\ref{APV_DDH}), which must be determined experimentally. The strategy to determine the spin-isospin structure of the HWI is to measure enough asymmetries for which we have theoretical predictions, to constrain and ultimately determine all of the 6 couplings.

An inherent problem in the experimental determination of the spin-isospin structure of the HWI is that asymmetries in calculable few-body systems are very small ($\sim 10^{-7} \rightarrow ~\sim 10^{-8}$) and difficult to measure. In contrast, nuclei can have large asymmetries that are easier to measure, but usually have large nuclear structure uncertainties, which make the measured results difficult to interpret in terms of fundamental degrees of freedom.
For a summary of previous measurements of PV asymmetries that constrain the HWI see~\cite{Depl,Haxton,NPDG}. V.~A.~Vesna et al.~\cite{Vesna} previously reported an upper limit: $|A_{PV}| < 12 \times 10^{-7}$ at $90\%$ confidence, using an approach similar to that of the present work.

The measurement presented here was made in a few body system for which the asymmetry has been calculated by Viviani {\it et al.}, using both the DDH framework and chiral EFT~\cite{Viviani1,Viviani2}. The first calculation uses the DDH potential for the weak interaction and a combination of the AV18/UIX potential~\cite{AV18,AV18-2,UIX} and the method of hyperspherical coordinates to describe the strong nucleon-nucleon interaction. In this framework, the asymmetry is given by~\cite{Viviani1}
\begin{eqnarray}\label{APV_DDH}
\nonumber
    A_{PV} &=& -0.185h^{1}_{\pi} - 0.038h^{0}_{\rho} + 0.023h^{1}_{\rho}  \\
           & & -0.001h^{2}_{\rho} -0.023h^{0}_{\omega} + 0.050h^{1}_{\omega}~.
\end{eqnarray}
Using chiral EFT, including contact terms, and one- and two-pion exchange terms, Viviani {\it et al.}, find~\cite{Viviani2}
\begin{eqnarray}\label{APV_EFT}
\nonumber
    A_{PV} &=& -0.137h^{1}_{\pi} - 0.049h^{0}_{\rho} + 0.015h^{1}_{\rho}  \\
           & & -0.0001h^{2}_{\rho} -0.023h^{0}_{\omega} + 0.024h^{1}_{\omega}~.
\end{eqnarray}
Using the DDH best values and ranges for the coupling constants (all of which are of order $10^{-7}$), the $\mathrm{n{^3}He}$ asymmetry is predicted to be $A_{PV} = -0.6^{+8.3}_{-10.7} \times 10^{-8}$ and $A_{PV} = 2.1^{+13.3}_{-10.6} \times 10^{-8}$ for the two calculations respectively. The chiral EFT result corresponds to a cutoff parameter of $\Lambda = 550~\mathrm{MeV}$. More recently, Gardner {\it et al.}~\cite{Gardner} calculate $A_{PV} \simeq -1.8\times 10^{-8}$, based on the large-$N_c$ framework~\cite{Schindler,Phillips}. Our result is a major step toward a complete experimental determination of the spin-isospin structure of the HWI, providing an important benchmark for theory.

\section{Description of the experiment}

The n$^{3}$He experiment ran at the Fundamental Neutron Physics Beamline (FnPB)~\cite{Fomin}, at Oak Ridge National Laboratory, at the Spallation Neutron Source, from December 2014 to December 2015. A brief overview of the n${^3}$He setup is given here. A detailed description of the  experiment can be found in~\cite{TGTCHAM,McCrea,Kabir,Coppola,Hayes}.

Intense $1~\mathrm{GeV}$ proton pulses from the SNS accelerator are produced at a rate of $60~\,\mathrm{pulses/sec}$. These protons interact with a mercury target to produce few $\mathrm{MeV}$ neutrons that are moderated in liquid hydrogen, at $\approx 20~\mathrm{K}$~\cite{Fomin}, to produce pulses of cold neutrons with a Maxwell-Boltzmann energy distribution. The pulsed nature of the beam, the neutron energy distribution at the moderator, and knowledge of the distance from the moderator to the detector ($17.5~\mathrm{meters}$) allowed accurate determination of the neutron energy at the detector using the neutron time-of-flight (TOF). Neutrons were guided from the moderator to the experiment by a supermirror neutron guide ~\cite{Fomin} with cross sectional area of $10\,\mathrm{cm}$ horizontal by $12\,\mathrm{cm}$ vertical. The guide has a curved section shortly after the moderator, thus preventing direct line of sight from the experiment to the moderator, reducing fast neutron and gamma backgrounds. In general, slow neutrons from one pulse will overlap with faster neutrons in the following pulses. To prevent these ``wrap-around'' neutrons, the neutron energy range in each pulse was restricted to be between $2~\mathrm{meV}$ and $9~\mathrm{meV}$, using a pair of TOF choppers~\cite{Fomin}. This reduced systematic effects and optimized statistics. The corresponding neutron fluence, after the polarizer and integrated over the selected neutron energy range and the spatial beam profile, was $1.8\times 10^{10}~\mathrm{n/s/MW}$~\cite{Tang}. The average delivered proton beam power varied from $0.7$ to $1.4~\mathrm{MW}$.
\begin{figure*}[t]
  \begin{center}
  \includegraphics[width=0.9\textwidth]{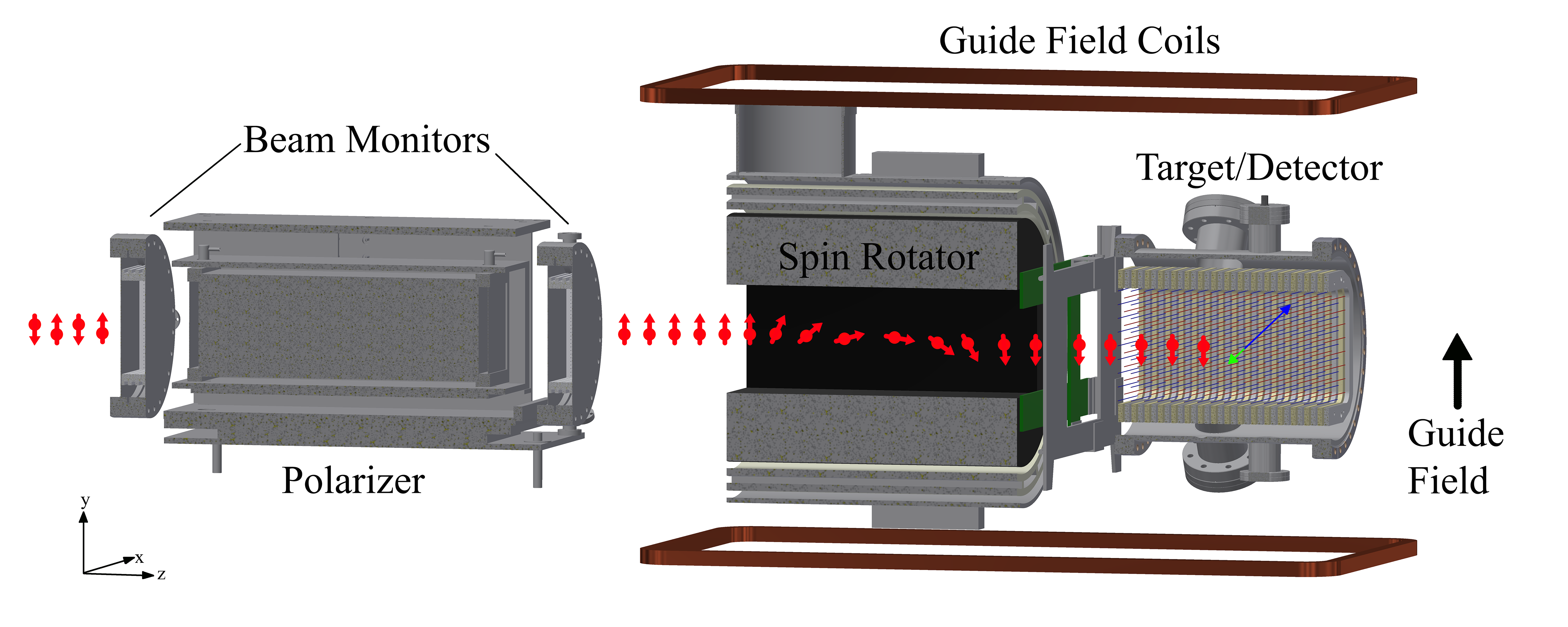}
    \caption{Illustration of the n3He apparatus. Neutrons enter from the left and travel in the $+\hat{z}$ direction. The beam monitor measures the relative neutron beam intensity and pulse shape. Neutrons exit the supermirror with spins aligned upward (along $+\hat{y}$). The RF spin rotator reverses the spin direction every other beam pulse.  Before entering the target the beam was collimated. Neutrons are captured in the target-detector chamber by $^{3}\mathrm{He}$, producing a proton and a triton per capture (the blue and green arrows respectively).}
    \label{fig:n3HeSetup}
  \end{center}
\end{figure*}

The experimental setup is illustrated in Fig.~\ref{fig:n3HeSetup}. In the beam direction ($+\hat{z}$), starting at the exit of the neutron guide, the experiment consisted of a beam monitor, a supermirror neutron polarizer (SMP)~\cite{Balascuta}, a resonant RF spin rotator (RFSR)~\cite{Hayes}, a 4-jaw collimator system, a holding field, and a target-detector ion chamber~\cite{TGTCHAM,McCrea}. A set of four race-track shaped magnetic field coils produced a 10 Gauss homogeneous field to hold the neutron polarization from the polarizer to the target. The holding field direction at the target was carefully aligned to the $+\hat{y}$ direction, the direction of neutron polarization after exiting the SMP. The beam monitor was a low efficiency $^3$He ionization chamber, absorbing only a few percent of the beam. It was used to monitor the relative neutron beam intensity and pulse shape to $10^{-4}$ fractional uncertainty in the intensity for a single pulse.

The RFSR reversed the neutron polarization for every other pulse. During production PV runs, the neutron beam was collimated to $8~\mathrm{cm}$ in $x$ (horizontal) by $10~\mathrm{cm}$ in $y$ (vertical). The neutrons captured in a combined target/detector wire chamber, filled with $^3\mathrm{He}$ gas at a pressure of $0.43$~atmospheres, at room temperature, absorbing the vast majority of beam neutrons in the selected energy range. The decay protons and tritons from the capture reaction ionized the $^3\mathrm{He}$ gas, and the charges were collected on the chamber wires and amplified to voltage signals. The target was separated into 144 wire cell volumes, defined by the 144 signal wires and the four HV wires surrounding each~\cite{TGTCHAM}. Wires were oriented perpendicular to the beam direction, either in horizontal or vertical orientation, depending on the measurement mode (see below). The charge collection properties of the chamber were simulated and tested and it was confirmed that there was negligible cross-talk between wire cells due to incomplete charge collection within a cell~\cite{TGTCHAM,McCrea}.

The measurement is based on the $^{3}\mathrm{He}(\mathrm{n},\mathrm{p})^{3}\mathrm{H}$ reaction ($Q = 764~\mathrm{keV}$). The energy of the final state particles is large compared to the center of mass energy of the initial state so that recoil effects are negligible and the ${^3}\mathrm{H}$ and $\mathrm{p}$ momenta are equal in magnitude and in opposite directions. Therefore, in the absence of parity violation, the cross-section is spherically symmetric. Radiative capture on $^3\mathrm{He}$ has a branching ratio of $10^{-8}$~\cite{WOLFS} and is negligible. The experiment's primary measurement was the directional asymmetry in the emission direction of the proton ($\hat{k}_p$), with respect to the neutron spin ($\hat{s}_n$). The corresponding single event differential cross section is given by
\begin{align}
\frac{d\sigma}{d\Omega} &= \left(\frac{d\sigma}{d\Omega}\right)_\mathrm{c}\left(1+A_{_{\mathrm{PV}}}\cos\theta_y + A_{_{\mathrm{PC}}}\cos\theta_x \right)\label{eq:DiffCrossSection}~.
\end{align}
Here, $\left(\frac{d\sigma}{d\Omega}\right)_\mathrm{c}$ is the unpolarized neutron capture cross-section, $A_{\mathrm{PV}}$ is the parity-violating (PV) asymmetry and $A_{\mathrm{PC}}$ is the parity-conserving (PC) asymmetry. The PV asymmetry is a result of the correlation $\hat{s}_n\cdot \hat{k}_p = \cos\theta_y$, while the PC asymmetry is a result of the correlation $ \left(\hat{s}_n \times \hat{k}_n\right)\cdot\hat{k}_p = \cos\theta_x$. For the definition of the PC correlation, we are generally using the coordinate system of Ohlsen and Keaton~\cite{Ohlsen}, but with the azimuthal angle $\phi$ measured from the spin axis $y$ to the scattering normal, $\vec{n} = \hat{k}_n \times \hat{k}_p$. In standard spherical coordinates $\cos\theta_y = \sin\theta\sin\phi$ and $\cos\theta_x = \sin\theta\cos\phi$.
\begin{figure*}[t]
  \begin{center}
  \includegraphics[width=0.95\textwidth]{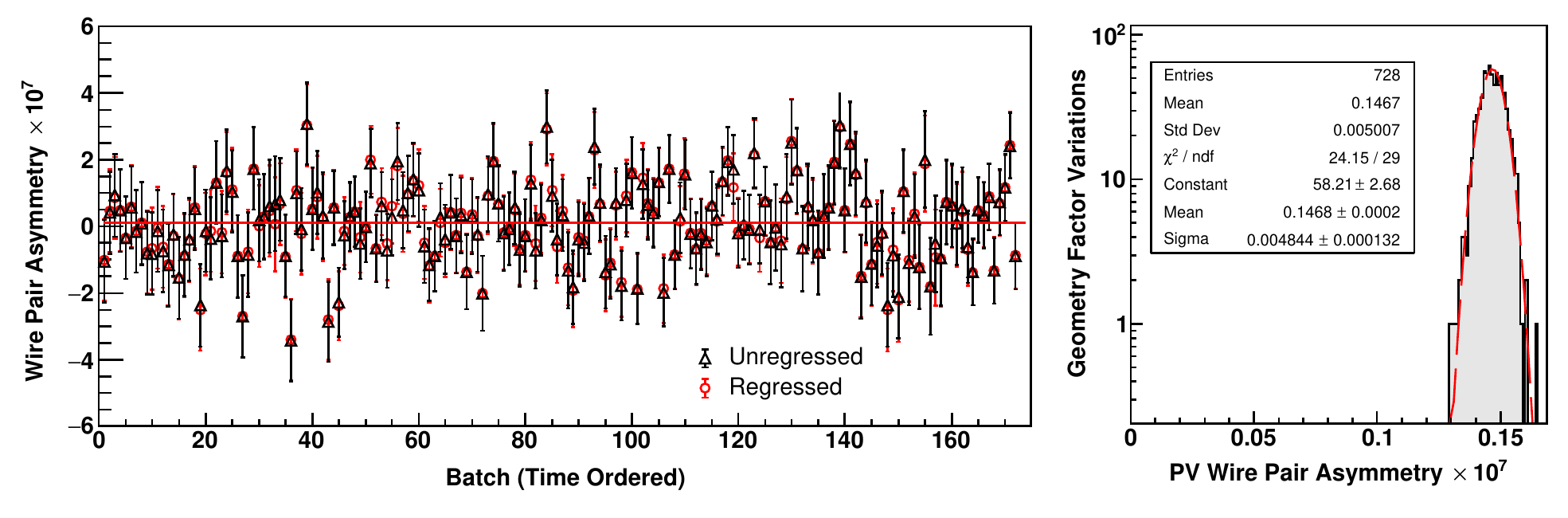}
    \caption{Left: Regressed and unregressed wire pair asymmetries by time ordered data batch number. Regression analysis of asymmetry versus random beam intensity fluctuations changed the central value by less than $0.04\times10^{-8}$. Right: Variation of PV wire pair asymmetries due to geometry factor uncertainty. The standard deviation sets the systematic error on the asymmetry due to uncertainty in the geometry factors (see text).}
    \label{fig:WPairAsymBatches}
  \end{center}
\end{figure*}

Referring to Fig.~\ref{fig:n3HeSetup}, since the beam polarization is purely transverse, along $\pm \hat{y}$ (with the beam momentum equal to $k_n\hat{z}$), the vector $\hat{s}_n\times \hat{k}_n$ lies along the $\pm\hat{x}$ direction. Therefore, when the neutron spin is reversed the sign of the correlation terms flips along the corresponding axis. The PV asymmetry was extracted by measuring the signal with wires in the upper and lower hemispheres ($\pm\hat{y}$), with the wires oriented horizontally ($\pm\hat{x}$) and the chamber centered on the beam. This orientation rejects the PC (left-right) asymmetry and we refer to it as up-down (UD) measurement mode in the remainder of the paper. The PC asymmetry was measured by rotating the chamber 90 degrees around the beam axis (counter-clockwise, looking in beam direction), so that the wires were oriented vertically, rejecting the PV asymmetry. We refer to this orientation as the left-right (LR) measurement mode.

\section{Data Analysis}

Performing an energy deposition and wire cell acceptance weighted average of $\cos\theta_x$ and $\cos\theta_y$ in Eqn.~\ref{eq:DiffCrossSection} yields an expression for the wire yields in terms of $A_{PV}$ and $A_{PC}$, given by
\begin{align}
Y^{\pm}_i &= Y^0_{i}\left(1\pm\epsilon P \left(A_{_{\mathrm{PV}}}G^{PV}_{i} + A_{_{\mathrm{PC}}}G^{PC}_{i}\right)\right)+p_i~.\label{eq:DetYield}
\end{align}
Here, the factors $\epsilon$ and $P$ represent the polarization reversal efficiency and beam polarization, respectively. For the $i^{th}$ wire cell, $Y^0_{i}$ is the spin-independent signal yield, $p_i$ is the signal pedestal, and $G^{PV}_{i}$, $G^{PC}_{i}$ are the so-called geometry factors, replacing $\cos\theta_y$ and $\cos\theta_x$ respectively. Asymmetries for each wire were formed for each pair of opposite polarization states (indicated by $\pm$ superscript)
\begin{equation}\label{eq:Asy1}
  A^{meas}_{i} = \frac{Y^{+}_i - Y^{-}_i}{Y^{+}_i + Y^{-}_i} ~.
\end{equation}
In each spin state, data were taken for $15.68~\mathrm{ms}$, separated into 49 TOF bins. Since the neutron polarization is nearly flat in the energy range selected by the choppers and the asymmetry is independent of neutron energy, the signal in each spin state was summed over the corresponding TOF range. Asymmetries were calculated either from single wire signals, according to Eqn.~\ref{eq:Asy1}, or for pairs of wires for which the horizontal plane bisecting the chamber formed the mirror image (i.e. wire pairs with opposite sign but equal magnitude geometry factors). The wire pair asymmetries were formed in two ways:
\begin{equation}\label{eq:Asy2}
  A^{meas}_{i} = \frac{1}{2}\left(\frac{Y_{u,i}^{+} - Y_{u,i}^{-}}{Y_{u,i}^{+} + Y_{u,i}^{-}} - \frac{Y_{d,i}^{+} - Y_{d,i}^{-}}{Y_{d,i}^{+} + Y_{d,i}^{-}}\right)~,
\end{equation}
and
\begin{equation}\label{eq:Asy3}
  A^{meas}_{i} = \frac{1}{2}\left(\frac{\frac{Y_{u,i}^{+}}{Y_{d,i}^{+}} - \frac{Y_{u,i}^{-}}{Y_{d,i}^{-}}}{\frac{Y_{u,i}^{+}}{Y_{d,i}^{+}} + \frac{Y_{u,i}^{-}}{Y_{d,i}^{-}}}\right)~.
\end{equation}
Where $u$ (up) and $d$ (down) refer to wires the same distance above and below the chamber mirror plane, respectively. The method corresponding to Eqn.~\ref{eq:Asy2} largely suppresses gain variations and any possible false asymmetry that couples to the gain, while method two (Eqn.~\ref{eq:Asy3}) suppresses beam fluctuations and the associated beam asymmetry. The pedestals were determined in weekly measurements and subtracted in the asymmetry denominator. The asymmetries calculated by all three methods were consistent with each other.

The measured asymmetries for each method were corrected for beam intensity asymmetries using linear regression with respect to the beam monitor data. The corresponding slopes were much larger for single wire asymmetries than they were for wire pair asymmetries, most of which had slopes at or below the few percent level.  The parity violating dataset was separated into 172 data batches containing about 185 runs each. Each run was about 7 minutes long and the total number of processed parity-violating data runs was 31854. The analysis produced 128 single wire asymmetries or 64 wire pair asymmetries, which were then combined in a least-squares fit to extract the physics asymmetries, as described below. The wire pair asymmetry (combined over all 64 pairs) vs. time ordered batch number is shown in Fig.~\ref{fig:WPairAsymBatches} (left).

The final analysis took into account two types of systematic effects: Those that multiply the asymmetry and those that add a correction. These are listed separately in Table~\ref{tbl:SYSEF}. The largest additive correction to the measured asymmetry is associated with a twist in the wire frame stack ($0$  to $20~\mathrm{mrad}$ front to back), that was observed during installation and carefully measured using survey equipment. The twist causes a correction, because it produces a wire rotation away from the horizontal, causing mixing between the LR and UD measurement modes, leading to the presence of PV and PC components in both sets of geometry factors.

To extract $A_{PV}$ and $A_{PC}$, the data were analyzed by a least-squares fit of the measured wire pair asymmetries from the UD and LR mode datasets, to the coupled set of equations
\begin{eqnarray}\label{eqn:LeastSq} \nonumber
  A^{meas}_{UD,i} & = & \epsilon P \left( A_{PV} G^{PV}_{UD,i} + A_{PC} G^{PC}_{UD,i} \right) \\
  A^{meas}_{LR,i} & = & \epsilon P \left( A_{PV} G^{PV}_{LR,i} + A_{PC} G^{PC}_{LR,i} \right) ~,
\end{eqnarray}
taking into account the correlation between wire asymmetries, which arises as a result of the long path length of the proton and triton through multiple wire cells. The correlations were obtained from measurement and verified with simulations. 
Neglecting the frame twist would remove the off-diagonal elements, $A_{PC} G^{PC}_{UD,i}$ and $A_{PV} G^{PV}_{LR,i}$, and produce the uncorrected asymmetries (not corrected for systematic effects) $A^{uc}_{PV} = (1.22 \pm .91~\mathrm{(Stat)})\times10^{-8}$ and $A^{uc}_{PC} = (-41.0 \pm 5.6~\mathrm{(Stat)})\times10^{-8}$. \newline

The geometry factors ($G^{PV}_{UD,i}$ and $G^{PC}_{UD,i}$ for UD mode and $G^{PC}_{LR,i}$ and $G^{PV}_{LR,i}$ for LR mode) were determined by performing simulations with the nominal beam collimation and chamber design geometry and then varying position and fill pressure parameters within their physical measurement uncertainties (e.g. survey) to minimize the difference between simulated and measured wire yields. This minimization process established the uncertainty on the geometry factors. The corresponding uncertainty in the asymmetry was determined by repeating the $\chi^2$-minimization of Eqns.~\ref{eqn:LeastSq} for each simulated set of geometry factors. For the PV asymmetry, the result is shown in Fig.~\ref{fig:WPairAsymBatches}(right).

A possible overall rotation of the wire frame stack with respect to the holding field would also mix the PV and PC asymmetries. The rotation angle was measured to be zero, with an uncertainty of $3~\mathrm{mrad}$, using field probes and survey equipment. The corresponding uncertainty in the PV asymmetry is $A_{PC} \times 3\times 10^{-3} \simeq 0.1\times 10^{-8}$, 7 times smaller than our statistical error.  A possible false asymmetry from the RFSR signal coupling to the front-end detector and DAQ electronics was measured during weekly beam-off runs. The averaged beam-off or pedestal asymmetry is $A_{ped} = (0.024 \pm 0.2)\times 10^{-8}$. The $^3$He target material produced extremely low background, being essentially insensitive to gamma background. The signal background from neutron capture induced $\beta$-decay in the target windows and other chamber materials was investigated using simulations and signal decay patterns in the chamber during beam-off periods; none were seen. Stern-Gerlach steering was evaluated based on the measured field gradient in the experiment holding field. The beam polarization and spin-flip efficiency were measured in dedicated runs~\cite{Hayes}. The final result, including statistical and all systematic error is
\begin{equation}\label{eq:FineAsy}
    A_{PV}=(1.58 \pm 0.97~\mathrm{(Stat)} \pm 0.24~\mathrm{(Sys)})\times 10^{-8}~.
\end{equation}

\begin{table*}
\caption{Systematic Corrections and Errors.\label{tbl:SYSEF}}
\begin{ruledtabular}
\begin{tabular}{cccc}
Additive Sources & Comment & Correction $[ppb]$ & Uncertainty $[ppb]$  \\ \hline
Frame Twist ($0$ to $20$ mrad) & compare simulation and data~\cite{TGTCHAM}  & $2.5$       & $0.2$   \\
Electronic false asymmetry     & measured~\cite{TGTCHAM}                     & $0.0$       & $2.0$   \\
Chamber field alignment        & compare simulation and data~\cite{TGTCHAM}  & $0.0$       & $1.3$\\
Mott-Schwinger scattering      & published calculation~\cite{MSCHW}          & $0.06$      & $0$ \\
Residual $^3$He Polarization   & calculation                                 & $< 0.06$    & $0$ \\
Background ($\beta$, $\gamma$) & simulation and calculation                  & $<< 0.1$    & $0$ \\
In-flight $\beta$-decay        & calculation~\cite{NPDG}                     & $<< 0.1$    & $0$ \\
Stern-Gerlach steering         & measurement and calculation ($\leq 2mG/cm$) & $<< 0.1$    & $0$ \\
\hline
Total                          &                                             & $2.6$       & $2.38$ \\
\hline \hline
Multiplicative Sources & comment & Correction & Uncertainty \\ \hline
Geometry factors               & compare simulation and data~\cite{TGTCHAM}  & $0.0$       & $0.5$ \\
Polarization                   & measurement~\cite{Hayes}                    & $0.936$     & $0.002$ \\
Spin-flip efficiency           & measurement~\cite{Hayes}                    & $0.998$     & $0.001$ \\
\hline
Total uncertainty              &                                             &             & $2.43$ \\
\end{tabular}
\end{ruledtabular}
\end{table*}

\section{Conclusion}

This result provides an important benchmark that extends our knowledge of the spin-isospin structure of the hadronic weak interaction, because the uncertainty in $A_{PV}$ is an order of magnitude smaller than the current theoretical reasonable ranges. The NPDGamma collaboration reported a measurement of the isovector pion coupling $h^{1}_{\pi} = (2.6 \pm 1.2)\times 10^{-7}$~\cite{NPDG}. If we insert this value into Eqn.~\ref{APV_DDH}, the contribution to $A_{PV}$ is $-4.9\times 10^{-8}$, indicating that there must be considerable cancellation between the $h^{1}_{\pi}$ term and heavy meson terms.

\begin{figure}
  \begin{center}
  \includegraphics[width=0.9\columnwidth]{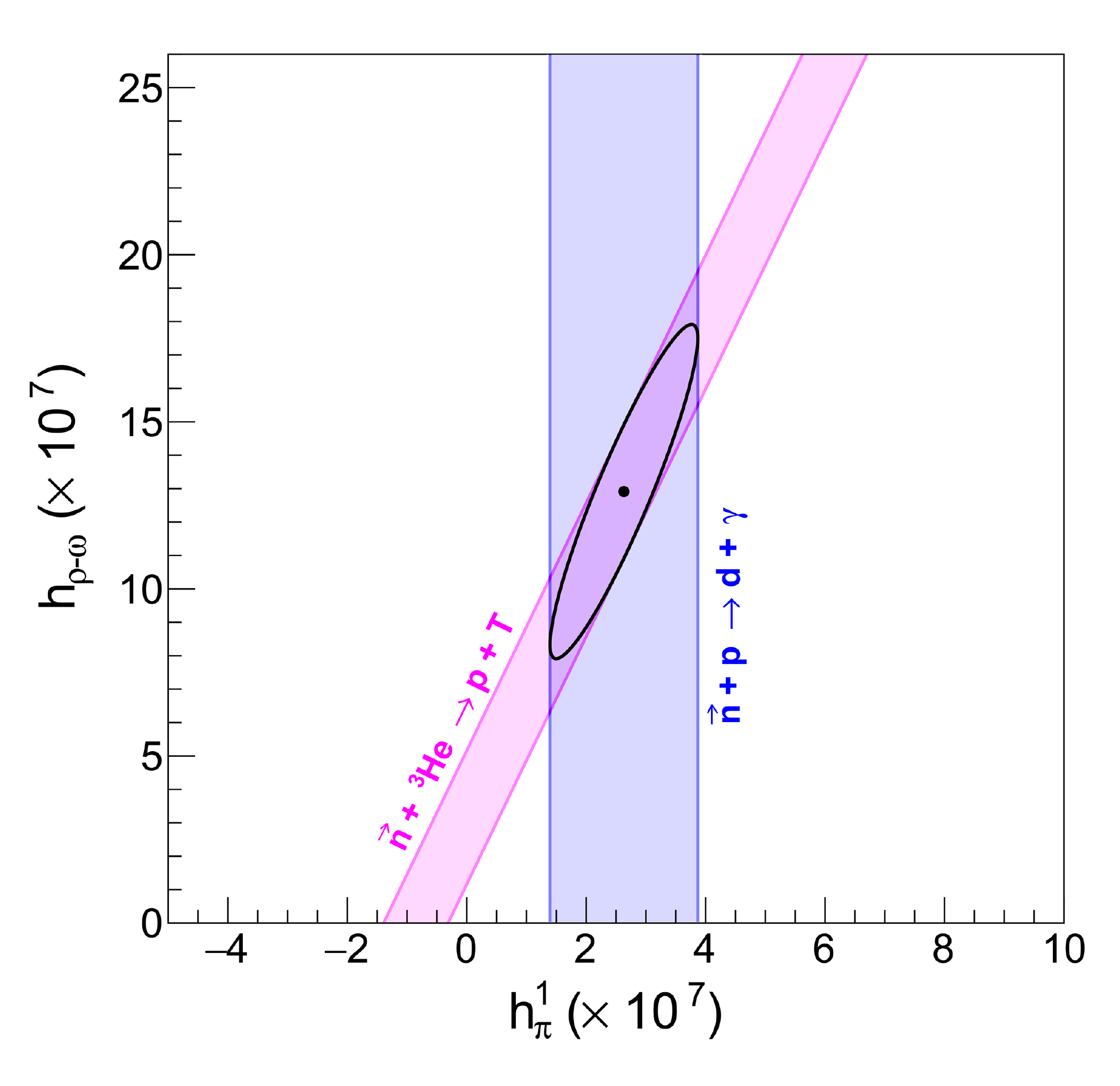}
    \caption{A least squares fit to the NPDGamma~\cite{NPDG} asymmetry
    and the n$^3$He asymmetry gives a constraint on a combination of heavy meson couplings, where
    $h_{\rho-\omega} \equiv h^{1}_{\omega} +0.46h^{1}_{\rho} - 0.46h^{0}_{\omega} - 0.76h^{0}_{\rho}  - 0.02h^{2}_{\rho} = \left(12.9 \pm 5.7\right)\times10^{-7}$}
    \label{fig:EXCLPlot}
  \end{center}
\end{figure}

When our result is combined with the NPDGamma asymmetry~\cite{NPDG} a constraint on a linear combination of heavy-meson couplings is obtained. These constraints are shown in Fig.~\ref{fig:EXCLPlot}. A least squares fit to the two asymmetries gives
\begin{eqnarray}\label{HEAVYC} \nonumber
  h_{\rho-\omega} &\equiv& h^{1}_{\omega} +0.46h^{1}_{\rho} - 0.46h^{0}_{\omega} - 0.76h^{0}_{\rho}  - 0.02h^{2}_{\rho} \\
                  &=& \left(12.9 \pm 5.7\right)\times10^{-7}~.
\end{eqnarray}

This analysis is possible because both reactions have been calculated with small model uncertainty, using the DDH potential model of the hadronic weak interaction. In order to improve our knowledge of the spin-isospin structure of the hadronic weak interaction additional measurements in few-body systems with small experimental uncertainties are required. Equally important are calculations of the asymmetries with small model uncertainties.

\section{Acknowledgements}

We gratefully acknowledge the support of the U.S. Department of Energy Office of Nuclear Physics through grant No. DE-FG02-03ER41258,
DE-AC05-00OR22725, DE-SC0008107 and DE-SC0014622, the US National Science Foundation award No: PHY-0855584, the Natural Sciences and Engineering Research Council of Canada (NSERC), and the Canadian Foundation for Innovation (CFI). This research used resources of the Spallation Neutron Source of Oak Ridge National Laboratory, a DOE Office of Science User Facility. We also thank Michele Viviani (INFN Pisa) for very fruitful theory discussions and Jack Thomison (ORNL) for his design support.

\bibliography{n3HePRL}

\end{document}